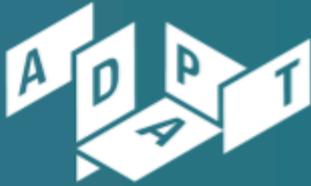

# ADAPT Centre Contribution on Implementation of the EU AI Act and Fundamental Right Protection

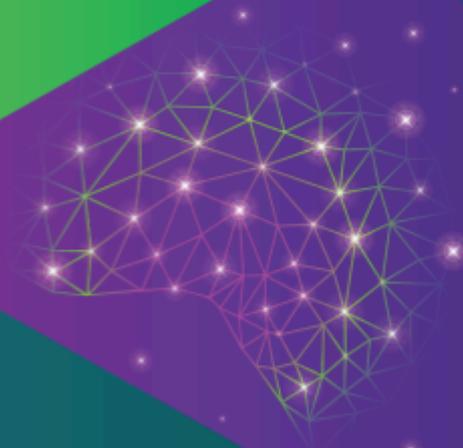

## Contributors:

Dave Lewis, TCD
Marta Lasek-Markey, TCD
Harshvardhan J. Pandit, DCU
Delaram Golpayegani, TCD
Darren McCabe, TCD
Louise McCormack, UoG
Joshua Hovsha, TCD
Deirdre Ahern, TCD
Arthit Suriyawongkul, TCD

# 1 Context

- The EU AI Act introduces a blanket protection of fundamental rights for specific applications of AI that it classifies as high-risk, which is implemented under the existing single market harmonised product certification mechanisms for health and safety protection, i.e. the New Legislative Framework.
- This protection of fundamental rights places many AI issues previously covered by voluntary trustworthy or ethical AI frameworks into a framework with independent and legally binding accountability for harmful characteristics of products grounded in the same human rights framework underpinning Union Law and many national laws.
- However, this major change in accountability also introduces many legal uncertainties on how AI providers and deployers can identify and manage risks to fundamental rights.
- Contrast this to the introduction of GDPR, which focussed on the protection of rights of privacy and data protection but benefitted from the development and employment of data protection principles under the data protection directive which had been in force beforehand. The protection of fundamental rights in AI systems however benefits from no such breakdown of principle, nor from prior deployment or compliance experience with such principles. This presents an extremely high level of legal uncertainty for providers and deployers of AI systems once the Act comes into force. The associated burden or chilling effects may fall disproportionately on public bodies wishing to deploy and reap the benefits of AI in high risk areas, and indigenous companies and especially SMEs that wish to market products into such applications.
- Public bodies, for example, face a new requirement to undertake a fundamental right impact assessment (FRIA) before considering the deployment AI in high risk applications (AIA Article 27(1)), but the acceptable form and content of such an assessment remains to be defined (AIA Article 27(5)), as does its role, if any, in the public procurement of such AI systems (AIA Article 62(3)(d)).
- While the AI Act does specify requirements for bodies such as the European Commission and the newly formed AI Office to develop guidance that will address some of the Act current legal uncertainties, it also relies on a complex network of competent authorities at national and union level to cooperate on regulatory learning to resolve these uncertainties.
- We maintain, however, that the relative immaturity of the state of the art in identifying, assessing and treating risks to fundamental rights from AI, combined with the rapidly advancing capabilities of the technology and the highly fluid modalities in how AI can be used will require extremely high levels of coordination in how relevant regulatory learnings are captured and shared.
- The dispersed and continually evolving nature of state-of-the-art expertise across various disciplines, government departments, public bodies, industry

- sectors, and civil society organisations necessitates the creation of effective and open multi-stakeholder mechanisms. These mechanisms are crucial for gathering, comparing, and synthesising new guidance and advanced understanding of technology, its risks to fundamental rights, and strategies for mitigating those risks.
- The broad scope and complex avenues by which AI may impact fundamental rights, and the ability of individuals to report such impacts, means that the regulatory learning mechanisms and multi-stakeholder deliberations on risks to and protections of fundamental rights must be conducted with an unprecedented ability to transparently communicate, update and explain their status and direction to the public, in order to build and maintain trust in AI innovation and use, especially in public service.
- Regulators and stakeholders must engage in rapid and effective regulatory learning and collaboratively develop a state-of-the-art understanding of risks to fundamental rights and their management. These insights should be transparently shared with the public. By making these processes accessible and treating them as public goods, they can also establish a gold standard to underpin voluntary codes of practice. This strategy not only aids AI value chain actors in building trust with customers and consumers but also promotes ethical standards and accountability across the industry. Moreover, it extends the benefits beyond the regulatory scope of the Act, fostering a collaborative environment and enhancing public confidence in AI technologies.

## 2 Mapping the Space of Current Legal Uncertainty

The above issues represent a complex and highly interconnected set of challenges for the implementation of the AI Act. It will require comprehensive communication and expert consensus built at a cross-EU level if the Act is to be successful in combining the protections needed to ensure public trust in AI while enabling competitive, effective and efficient value chains for AI systems, data and components across the single market.

Competent authorities in Ireland must, therefore:
1. Engage effectively with the EC and AI Office, European standards organisations involved in developing harmonised standards and with peer authorities in other member states to communicate and ensure timely progress on items requiring legal certainty that reflect national priorities and the implementation of the national AI strategy;
2. Identify and adequately resource mechanisms within Ireland that can contribute to resolving legal uncertainty within its jurisdiction and also offer influential regulatory learning outcomes to the EC and peer competent authorities.

Below we present a problem space for potential legal uncertainties in implementing the Act as a guide to positioning and prioritising instruments for regulatory guidance and regulatory learning.

The problem space for addressing legal uncertainty under the AI Act, can be mapped along the following three axes:

- **Protections:** These may include protection of health and safety as established in existing harmonised legislation under the new legislative framework, but also the novel protections for the full range of fundamental rights as laid out in the EU Charter of Fundamental Rights, including protection for equality, democracy, the rule of law and the environment.
- **Types of AI Systems defined in the AI Act,** including prohibited systems, high-risk AI (HRAI) systems as defined in Annex I and in Annex III, and high-risk AI systems declared as non-high risk
- **Use cases representing different AI value chains**, ranging from the central case of high risk AI provision, General Purpose AI (GPAI) model provision, AI provision under public procurement regime, provisional AI system assessment under regulatory sandboxes or human trials, use cases involving substantial change to the deployed AI system and use cases where unanticipated risks materialise after deployment as identified in incident reporting and reporting by impacted stakeholder.

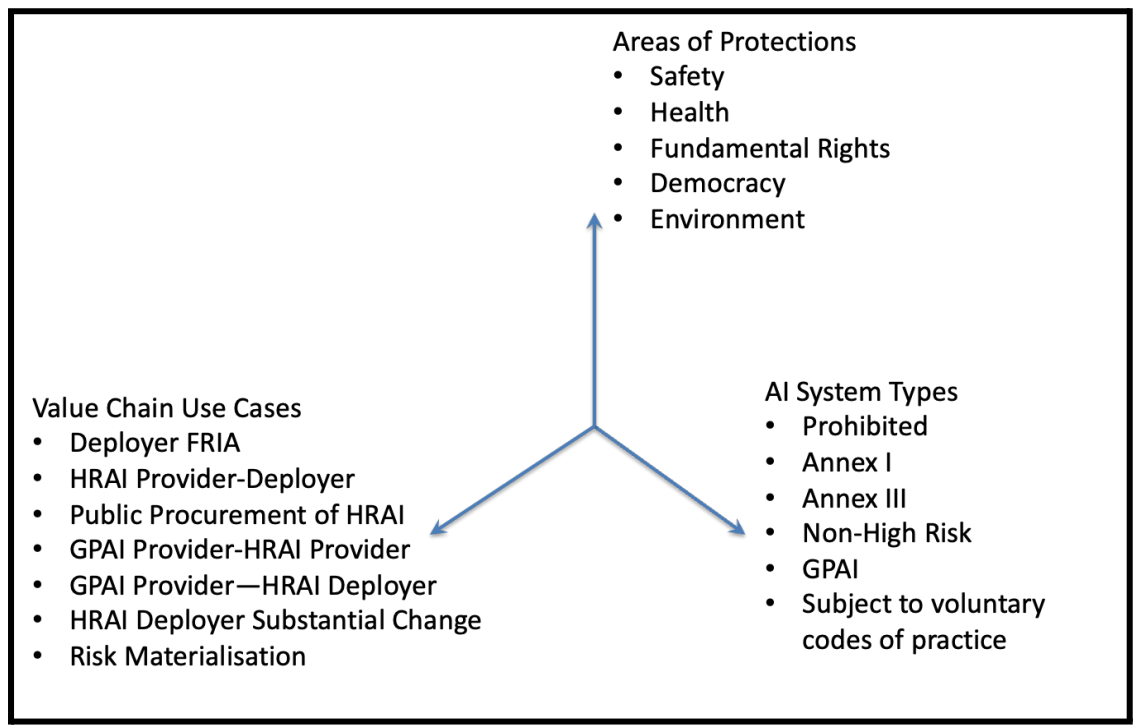

Figure 1: Major axes defining the uncertainty space of the AI Act within which regulatory elaboration and learning mechanisms could be placed.

## 2.1 Protections related to Fundamental Rights

When referring to fundamental rights protection in AI governance, the EU AI Act has largely followed the language of the EU Charter of Fundamental Rights (CFR). The Preamble to the AI Act, in Recital 48, specifically mentions 17 rights, which translate into 23 Charter Articles, as presented in the table below:

| | |
|---|---|
| Right to human dignity | Art 1 CFR |
| Respect for private and family life | Art 7 CFR |
| Protection of personal data | Art 8 CFR |
| Freedom of expression and information | Art 11 CFR |
| Freedom of assembly and association | Art 12 CFR |
| Non-discrimination | Art 21 CFR |
| Right to education | Art 14 CFR |
| Consumer protection | Art 38 CFR |
| Workers' rights | Arts 15, 27, 28, 29, 30, 31 and 32 CFR |
| Rights of persons with disabilities | Art 26 CFR |
| Gender equality | Art 23 CFR |
| Intellectual property rights | Art 17(1) CFR |
| Right to an effective remedy and to a fair trial | Art 47 CFR |
| Right of defence and the presumption of innocence | Art 48 CFR |
| Right to good administration | Art 41 CFR |
| Children's rights | Art 24 |
| Environmental protection | Art 37 CFR |

However, it appears that other fundamental rights protected by the Charter, which have not been explicitly mentioned in the AI Act, may also be impacted by the new legislation. For example, the prohibition of using biometric categorisation systems that categorise individually natural persons based on their biometric data to deduce their religious or philosophical beliefs is grounded in the right to freedom of thought, conscience and religion enshrined in Article 10 CFR, which is not explicitly mentioned in Recital 48 to the AI Act. Preliminary research into this area suggests that the AI Act may potentially impact all the substantive rights included in 50 Articles of the Charter (excluding the so-called horizontal provisions laid down in Articles 51-54), whether directly or indirectly. Importantly, in accordance with Article 52(1) of the EU Charter of Fundamental Rights, any limitations on the exercise of fundamental rights must not only be made subject to the principle of proportionality, but it must also respect the essence of those rights.

The potential impact of fundamental rights on the AI governance in the EU is particularly visible in Article 27 of the AI Act, which introduces the requirement to carry out an *ex ante* fundamental rights impact assessment (FRIA), which will be mandatory for most deployers of high-risk AI systems. While a template for a questionnaire, including through an automated tool, will be developed by the future AI Office, the implementation of Article 27 is likely to pose significant challenges in practice. Even though impact assessments related to various fundamental rights are a commonly used methodology, including in EU Law, the existing methodologies to date have focused on assessing the impact on one isolated right or a group of related rights (e.g. environmental impact assessment, data protection impact assessment, gender, disability or equality impact assessment). Conversely, by introducing a duty to conduct a fundamental rights impact assessment, the EU is laying down a requirement of a comprehensive review, presumably taking account of all the fundamental rights that may potentially be affected in varied contexts and impacting differently situated individuals by a given AI system.

Another challenge in the implementation of the AI Act is going to be the extent of the impact of different fundamental rights on the deployment of AI systems. The legal basis of the AI Act is twofold: 1) Article 16 TFEU, which lays down the EU's competence concerning the protection of the right to privacy, and 2) Article 114, which concerns harmonisation for the internal market. Thus, the protection of the fundamental right to privacy will be a priority concern for the implementation of the AI Act. The same, however, cannot be said about the remainder of Charter rights, which have been positioned merely as overriding reasons of public interest, justifying restrictions to the free movement on the internal market. It seems unclear how AI innovation is to be balanced against fundamental rights concerns under the AI Act. It appears that – except for the right to privacy – every measure taken to protect fundamental rights will have to be balanced against the overarching objective of facilitating the free movement of AI systems in accordance with the principle of proportionality.    This gives an idea of how complex and uncharted this territory is for competent authorities, the EU institutions, and by extension, firms falling within the AI Act's relevant provisions.

## 2.2 AI System Types:
The provisions of the AI Act are targeted at specific types of AI systems defined in terms of the application area the system intends to address, with large AI systems not intended for specific application addressed separately as these present different forms of potential legal uncertainty.

| Class of AI system | Potential uncertainties | Body to direct questions on uncertainty |
|---|---|---|
| Prohibited Applications (Art.5) | What qualifies as a 'subliminal technique' under Art.5.1(a)? What qualifies as a 'purposefully manipulative or deceptive technique' under Art.5.1(a)? | |
| High Risk AI Systems subject to Union Harmonised Legislation Annex I (Art 6(1)) | What fundamental rights are relevant/not relevant to a given harmonised legislation scope? How does the FR for life (CFR Art 2) and integrity of the person (CFR Art 3) relate to the interpretation of health and safety protections, e.g. in an FRIA? | Initially notified bodies (they all need external conformity assessment under the AI act?) and then the relevant market surveillance authority |
| AI Systems listed in Annex III (Art 6(2)) | AI systems included in Annex III are potentially classified as high-risk, subject to complex and overlapping conditions and exceptions outlined in Articles 6(3) and 6(4). | The Commission, with consultation from the European AI Board (EAIB), will provide guidelines specifying the practical implementation no later than 2 Feb 2026.<br><br>To resolve uncertainty, the Commission also has power to amend Art 6(3) subpara 2 and Annex III (Arts 6(6), 6(7), 6(8), 7). |
| AI Systems listed in Annex III self-assessed to be not high risk (Art 6(4)) | What regime will be in place for checking if such self-assessment correctly interprets risk levels? | - |
| AI Systems listed in Annex III that does not pose a significant risk of harm to the health, safety or fundamental rights | What regime will be in place for checking if such self-assessment correctly interprets the stated derogations in Art 6(3) ? | - |
| of natural persons (Art 6(3)) | | |
| AI Systems listed in Annex III that performs profiling of natural persons (Art 6(3) subpara 3) | What qualifies as profiling, especially if integrated into more complex AI processing and inference ? | - |

| Regulated Sector | Harmonised legislation | Responsible Irish Body |
|---|---|---|
| machinery | 2006/42/EC | Health and Safety Authority |
| toys | 2009/48/EC | Competition and Consumer Protection Commission |
| recreational/personal watercraft | 2013/53/EU | Dept of Transport |
| lifts | 2014/33/EU | Health and Safety Authority |
| explosive gases | 2014/34/EU | Health and Safety Authority |
| radio equipment harmonised legislation | 2014/53/EU | ComReg |
| pressure equipment harmonised legislation | 2014/68/EU | Health and Safety Authority |
| cableway installation harmonised legislation | 2016/424 | Commission for Railway Regulation |
| personal protective equipment harmonised legislation | 2016/425 | Health and Safety Authority & Competition and Consumer Protection Commission |
| burning gaseous fuels harmonised legislation | 2016/426 | Health and Safety Authority & Competition and Consumer Protection Commission |
| medical devices harmonised legislation | 2017/745 | Health Products Regulatory Authority |
| in vitro diagnostic medical devices harmonised legislation | 2017/746 | Health Products Regulatory Authority |
| civil aviation harmonised legislation | 300/2008 | Irish Aviation Authority |
| two- or three-wheel vehicles and quadricycles harmonised legislation | 168/2013 | Under consideration |
| agricultural and forestry vehicles harmonised legislation | 167/2013 | Minister for Agriculture, Food and the Marine |
| marine equipment harmonised legislation | 2014/90 | Department of Transport. Marine Survey Office |
| rail systems harmonised legislation | 2016/797 | Commission for Railway Regulation |
| motor vehicles and their trailers and components harmonised legislation | 2018/858 | Road Safety Authority of Ireland |

| civil aviation safety harmonised legislation | 2018/1139 | Irish Aviation Authority |

## 2.3 AI Value Chain Use Case

Developing structures and support for effective and efficient implementation of the AI Act requires careful consideration of interactions that need to occur between
Mapping and Engaging AI Value Chain Actors:
- Identify Key Stakeholders: Map all relevant stakeholders across the AI value chain, including:
  - Different configurations of AI value chain actors,
  - Notified bodies and market surveillance authorities serving that chain and
  - The stakeholders whose health, safety and fundamental rights are potentially impacted by AI systems and who must interact relevant value chains and regulatory actors

AI Value chain interactions addressed by the act include:
- High-risk AI provider interacting with an AI deployer
- High-risk AI provider interacting with a public sector AI deployer through a public procurement process
- GPAI provider interacting with a high-risk AI provider
- GPAI provider interacting with a high-risk AI deployer
- Supplier of AI systems, tools, services, components, or processes interacting with an high-risk AI provider

For high-risk AI systems that require certification by a notified body or undertaking self-assessment should differentiate between:
- Support existing certification assessment bodies in Ireland that aim to act as notified bodies under the AI Act, both in obtaining and maintaining notified status and in attracting parties seeking certification from them.
- Determining the ambition to establish new notified bodies in Ireland, and then support those bodies
- Support for enterprises based in Ireland in preparing to seek certification from a non-Irish notified bodies (e.g. when an appropriate body in not established in Ireland)

Considering
- Support for enterprises based in Ireland in offering GPAI to a high risk provider or deployer in the Union
- Support for enterprises based in Ireland undertaking a contract to supply AI system, tools, services, components, or processes to a high risk AI provider in the Union

# 3. Consideration for configuration of National Competent Authorities

*For national implementation of the Act, different approaches to the designation of competent authorities could be considered, ranging from a centralised model to a more distributed, sector-based approach. Selecting an approach will likely involve trade-offs. For example, a distributed approach may provide better access to sectoral expertise, but may pose coordination challenges.*

***What considerations should the Department have regard to when devising the configuration of national competent authorities for implementation?***

Taking a centralised approach to configuring competent authorities in the form of an AI Authority (or Digital Regulator as Italy is doing) may have advantages. First, it would allow for cross-cutting expertise and resources to be built up and harnessed in one entity who would be a credible super-regulator for AI. Secondly, it would create a visible, streamlined one-stop shop approach that would appeal to firms of all sizes rather than having to approach a variety of regulatory bodies with the potential for inconsistencies of approach. This approach would be pro-innovation and regulatory learning. A single AI regulator is seen in Spain and is likely to emerge in the UK. An alternative would be to have an efficient formal cooperation forum between digital regulators for regulatory learning and consistency. It would also make sense to harness lessons learned from the Central Bank's planned roll-out of a regulatory sandbox.

## 3.1 Implementing Fundamental Rights Protections:

On the basis that the state of the art in mechanisms and measures for protecting fundamental rights in AI product life cycles is relatively immature and experience in conformity testing against fundamental rights is extremely rare, implementation of the AI Act must prioritise efficient and timely development of applicable knowledge, accessible resources and guidance, deployable expertise, key competencies and channels for stakeholder engagement.

The AI Act, as a Regulation, will have direct effect. Most Charter rights do not have direct effect, except the ones that are already protected in other legislation with direct effect (e.g. discrimination or data protection). A prerequisite therefore for legal certainty for the implementation of the Act is that the value chain, regulatory and stakeholder representative actors in any high-risk sector must have understanding of the relevant EUand national legislation that does have a direct effect in protecting fundamental rights. Further, in theory, this seems to imply that in future litigation under the AI Act will override those Charter rights that don't have direct effect, eg. the right to freedom of expression.This would seem to place a lot of interpretive power on the protection of some fundamental rights within the remit of current systems for product regulation.  It is not clear if this is an explicit intent of the Act or even if this a desirable outcome, raising broader l questions about the role of the Act in the protection of fundamental rights in general.

**Recommendations:**
- The National Competent Authorities must liaise with the EC, AI Office and European AI Board (EAIB) to seek guidance on the set of Union law that have a direct effect on the protection of fundamental rights and their relevant application to the different areas of AI covered by the Act, e.g. prohibited, high-risk, non-high risk, GPAI
- The National Competent Authorities should analyse the Union-level guidance on relevant legislation with a direct effect on protection of fundamental rights and map that into the implementation of that Union legislation in national legislation. They should also assess which other national legislation offers direct effect in the protection of fundamental rights in the sectors covered by the AI Act. This should be undertaken in collaboration with the Irish Human Rights and Equality Commission. Cooperation between relevant government departments is also recommended in undertaking this analysis, such as the Workplace Relations Commission, Department of Social Protection, Health Insurance Authority, International Protection Office and the Data Protection Commission. As this is a potentially large legal analytical task, we suggest the following sectors are prioritised:
    - Employee recruitment, promotion, evaluation & termination of employment,
    - Eligibility for social security payments, grants & benefits
    - Pricing of life or health insurance
    - Examination of applications for asylum, visa or residence permits
- The National Competent Authorities must liaise with the EC, AI Office and the EAIB to seek clarity on the intended scope of the Act in protecting fundamental rights, especially for rights that are not protected by other Union legislation or national legislation with direct effect. As part of this, the mechanism for resolving possible conflict between protections for different fundamental rights need to be clarified, in particular the role of the principle of proportionality in this.

## 3.2 Implementing National Responses to EU-level updates:

Integrated legal and technical guidance is required to address the legal uncertainties currently presented by the AI Act in the near and medium term. The Act already indicates where specific guidance and resources will be developed by the AI Office; by the EC through implementing and delegated acts or common specifications; by European Standards Organisation for harmonised standards, and by the EU AI Board through ongoing guidance and opinions.

**Recommendations:**
- National Competent Authorities should ensure that the likely wide range of implementation guidance is clearly contextualised for and communicated to local value chain, regulatory and stakeholder actors in a timely fashion, ideally through a single, searchable communication portal. Therefore, to prioritise clear and consistent communication during a wide and complex phase of regulatory learning, fragmentation of communication functions across different authorities should be avoided.
- National Competent Authorities should liaise closely with Union level bodies to assemble a running schedule of when guidance, standards, opinions and implementing or delegated acts will be appearing, including the timing of opportunities to be consulted on their preparation.
- The AI Advisory Council, the Enterprise Digital Advisory Forum, and the GovTech Delivery Board should provide oversight and be able to make recommendations on the form, quality and timing of communications in relation to the implementation of the AI Act by National Competent Authorities.

## 3.3 Regulatory Learning mechanisms: Sandboxes, Testing in real-world conditions, and Incident reporting:

The current high level of legal uncertainty associated with fundamental rights protections and the rapid development of AI technology and its ability to impact those rights points to the critical importance of making effective and responsive use of the regulatory learning mechanisms in the Act, namely regulatory sandboxes, testing in real world conditions, and sharing of information on serious incidents. Uncertainties make it difficult to predict the optimal priorities for investment in such regulatory learning mechanisms.

**Recommendations:**
- The National Competent Authorities should liaise closely with the EC, AI Office and EAIB on the development of guidance for the implementation of regulatory learning mechanisms, contribute and encourage the open and timely sharing of learnings from the implementation of such mechanisms between Member States. The scope and focus of past and active mechanisms in other Member States should be closely monitored to ensure Irish planning and investment in such mechanisms does not needlessly duplicate those elsewhere.
- The National Competent Authorities should recognise that a single learning mechanism (i.e. a single sandbox) will be insufficient to address the wide range of uncertainties facing the implementation of the Act, and therefore an agile approach to establishing, operating and retiring multiple regulatory learning mechanisms should be adopted. A rolling strategic plan for managing the lifecycle of such mechanisms should be established, adopting a paradigm of a *regulatory learning mechanism factory* that aims to continuously improve the learning productivity, and minimise the lifecycle cost, of each mechanism (including the use of data sharing mechanisms - see section 4). We suggest that such mechanisms be positioned to serve as spaces at the intersection of AI system types and AI value chain types outlined in section 2, in order to best convene the relevant sectoral expertise from regulatory, value chain and stakeholder actors. For example, a sandbox to address the public procurement of a system employing AI in higher education and vocational training could be convened by DFHERIS, involving the HEI, education software providers and stakeholder groups including staff and student unions, professional accreditation bodies.
- The National Competent Authorities, in collaboration with other Member States, should develop effective mechanisms, including guidelines, template, and a common incident taxonomy, for collecting, sharing, and analysing serious incident reports. As informed by best practices from more mature cybersecurity incidents and vulnerabilities response communities.

# 4. Synergies with Other Digital Regulation
*The EU has adopted a series of Regulations in recent years designed to protect consumers, strengthen the internal market, and ensure that the EU remains at the forefront of innovation and the adoption of advanced technologies.*

***Are there potential synergies between the implementation of AI Act and the implementation of other EU Regulations applying to Digital markets, services, and infrastructure?***

## 4.1 GDPR
For AI Deployers, synergies may be possible between the need to undertake a Data Protection Impact Assessment (DPIA) under GDPR and the broader based requirement for a FRIA if planning to deploy an AI system in a high risk application. GDPR has a structured process for undertaking a DPIA: 1) assess need for DPIA, 2) carry out the DPIA, 3) determine whether processing should take place. We can express the FRIA process in a similar manner:
1. assess need for FRIA - which will be based on high risk categorisation;
2. carry out the FRIA - where we will need to identify what are inputs, and what the outputs will be e.g. analysis of the impact on specific fundamental rights, the assessment of the level of risk and its acceptability, tradeoff consideration between different risks and the expect benefits of the intended purpose of the AI system; and then
3. what should the FRIA lead to e.g. a halt to the intended use of AI, further legal and stakeholder consultation to consider viable treatments for unacceptable levels of residual risk and detailed requirements for building and procurement of the AI system.

Currently, we note that there is some variation in the formats of DPIA suggested and accepted by different member state GDPR supervisory authorities. While under GDPR such variation only impacts organisations with major data processing activities in multiple member states, for FRIA national variations in recommended/accepted formats may undermine the goal of fundamental rights protections that can be implemented in AI products that can be deployed seamlessly across the single market.

**Recommendations:**
- The National Competent Authorities should liaise closely with the DPC to ensure efficient alignment of FRIA and DPAI processes and guidelines, especially to minimise the requirements for duplication in bodies undertaking the assessments
- The National Competent Authorities should liaise closely with the AI Office to ensure strong normalisation of FRIA guidelines (Art 27.5) and with the European AI Board (EAIB) to ensure good coordination and interoperability on FRIA between market surveillance bodies, especially for the specific high-risk sectors.

## 4.2 Data Governance Act (DGA)

The DGA offers a legal grounding for organising data sharing for non-profit purposes. Given the need to rapidly develop guidance, benchmarks and measurement methodologies in areas of fundamental rights protections in the face of relatively immature state of the art, using DGA to facilitate the rapid and widespread sharing of information and data on FR risk assessment and mitigation testing.

**Recommendations:**
- National Competent Authorities, in coordination with the DPC, should liaise with the AI Office and the EAIB to seek best practice on employing DGA measures to support controlled data sharing for regulatory learning to support AI Act mechanisms, including sandboxes (Art 57, 58) and user testing in real world consideration (Art 60), including for the appropriate protection of shared personal data (Art 59) and consent of test subjects (Art 61). Such liaison should also address the use of or interaction with existing data space infrastructure for regulatory learning mechanisms, including the European Health Data Space[1] and the Public Procurement Data Space[2].
- Consideration should be given to establishing altruistic data sharing bodies in Ireland to support public interest interaction and exchange of measurement methodologies and benchmarking data between relevant: competent authorities; public sector actors; their private sector supplier value chains; the stakeholder groups most vulnerable to fundamental rights risks, e.g. patients, employees, those accessing public benefits/services; and groups that represent them, such as IPOSSE, the Irish Council for Civil Liberties and labour unions.

## 4.3 General Product Safety Act and the Product Liability Directive

These new legal instruments expand the scope of some product liability to cover impacts from AI. A further EU AI liability directive is also under development. These legal instruments may interact with the AI value chain in new ways that need to be reflected in contracts with upstream suppliers. There is already an upstream contractual requirement for high risk AI providers and any suppliers of AI systems, tools, services, components, or processes (Art.25.4) as well as on upstream information provision from GPAI provider (Art.53).

---

[1] https://health.ec.europa.eu/ehealth-digital-health-and-care/european-health-data-space_en

[2] https://single-market-economy.ec.europa.eu/single-market/public-procurement/digital-procurement/public-procurement-data-space-ppds_en

**Recommendations:**
- National Competent Authorities should liaise closely with the AI Office and EAIB on central guidance provided on supplier-high risk AI providers contract templates and information provision from GPAI providers.
- Consider adapting this guidance to provide templates and guidance for third party suppliers and GPAI provision contracts that are grounded in common law familiar to importers or distributors from North America.

## 5. Excellence in AI Regulation

*Harnessing Digital - The Digital Ireland Framework[3] establishes the goal for Ireland to be a digital leader at the heart of European and global digital developments. In support of this goal, Ireland is a member of the D9+ Group, an informal alliance of Digital Ministers from the digital frontrunner EU Member States. It also calls for Ireland to be a "centre of regulatory excellence" in Europe. The AI Act will set out a requirement to promote innovation, having regard to SMEs, including start-ups, that are providers or deployers of AI systems.*

**How can Ireland's implementation of the AI Act bolster Ireland's position as a leading Digital Economy, increasing investment and accelerating innovation in AI? What would excellence in AI regulation look like?**

- Dimension 1 Digital Transformation of Business
    - 1.1 Digitalisation of Enterprise: AI has major potential for improving economic productivity, but its uptake may be impeded by uncertainty about the regulatory obligations and associated legal and reputation risks, especially by SMEs which are less well resourced to resolve these legal uncertainties. By focussing on reducing these uncertainties and improving communication, our recommendations aim to ensure implementation of the AI Act minimises any chilling effect on AI adoption.
    - 1.2 Ireland as a location for leading digital enterprises: By focussing mechanisms for regulatory learning in Ireland on the interactions that occur across value chains, our recommendations aim to make Ireland attractive as a location where issues of contractual liability, fundamental rights risk assessment and generation of upstream

---
[3] https://www.gov.ie/en/publication/adf42-harnessing-digital-the-digital-ireland-framework/

- technical documentation are well supported. This builds on Ireland's existing leading position as a host location to many US firms that will operate as GPAI providers or as third party system/tool/data/process providers to AI providers in the European single market.
  - 1.3 Broader Economic Digital Dividends Dimension: By highlighting the relative immaturity of the state of the art in protection of fundamental rights compared to prior health and safety and personal data protection, our recommendations highlight the high potential to leverage the research and innovation expertise in Ireland in the areas of trustworthy and ethical AI and data governance. Close collaboration between National Competent Authorities and Ireland's research centres, technology centres and innovation hubs offers strong opportunities for mutual benefits. This includes transferring research results and regtech innovations between these centres and the regulatory mechanisms for the AI Act operating in Ireland such as regulatory sandboxes and real-world testing. . In particular, the expertise available in research centres for engaging in human-facing assessments and undertaking engaged research with relevant societal stakeholders may dramatically reduce the lead time in identifying and resolving uncertainties in managing risks to fundamental rights. Our recommendation also contributes to participation by local enterprises in public procurement of high-risk AI systems. This will help reduce barriers for enterprises exporting AI systems or related components to downstream customers across the single market. Finally, alignment with national R&I activities will open opportunities for enterprises to collaborate with researchers to access EU-funding aimed at improving the digital single market adoption of AI in compliance with the AI Act and to meet goals of the twin digital and green transition.
- 2 - Digital Infrastructure Dimension: The primary impact of the AI Act on digital infrastructure is the need for AI products to process and be monitored for appropriate levels of cyber security resilience., This is a requirement for high risk AI systems (AIA article 15) and for digital products in which they are contained under the new EU Cyber Resilience Act. Cyber security is also central to underpinning protections for many other rights, especially those related to personal data and its processing by AI.
- 3 - Skills Dimension: Our recommendations support several aspects of the skills dimension of the Digital Ireland Framework. Protection of the right to decent employment is one of the rights protected under the Charter as is protection against discrimination in both employment and education, with the impact of AI-based decision making in these subject to high risk FRIA. This provides a formal legal grounding for deliberating on the impact of adopting AI in the workplace and in education. Further, the complex needs of the regulatory process is itself part of the digital transition, and one that in Europe is not likely to diminish. Therefore the development of regulatory learning mechanisms in Ireland may also serve to catalyse the demand for and the provision of skills in AI regulation and governance. The strength both in the existing governance, risk and compliance sector in Ireland and the HEI research in this area indicate the potential for developing a skills hub in AI regulation. Our recommendation of prioritising regulatory learning for public sector procurement of AI points to the potential for public sector leadership in developing workforce skills in governance, risk and compliance roles across all sectors.
- 4 - Digitalisation of Public Services Dimension: our recommendation suggest prioritising regulatory learning mechanisms for public procurement of AI, as a focus for accelerating and maximising the propagation of learning on fundamental rights between the public sector, its commercial supplier and its sector-relevant societal stakeholders.

## 6. Alignment with the Objectives of the AI strategy for Ireland

*AI - Here for Good: National Artificial Intelligence Strategy for Ireland[4] sets out how Ireland can be an international leader in using AI to benefit our economy and society, through a people-centred, ethical approach to its development, adoption, and use. In recognition of the wide-ranging effect AI will have on our lives, this Strategy considers AI from several perspectives: Building public trust in AI; Leveraging AI for economic and societal benefit; and Enablers for AI.*

**How can Ireland's implementation of the AI Act drive support and accelerate progress from each of these perspectives while meeting our regulatory obligations?**

**The following expand how the above recommendations align with and can support actions proposed under the different Strands of the National Ai Strategy**
- **Strand 1: AI and Society:** Our recommendations focus on the need to ground activities on engaging the public on AI and building trust in the first instance in clear and consolidated communication about the guidance, decisions, standards and specifications that are required to be developed at a Union level, but with careful explanation and contextualisation of this into the Irish context. A core message to build here is that the AI Act in essence places AI more accountability into the remit of existing Union legislation, including harmonised product regulation, GDPR and the charter of fundamental rights. A core part of the trust building communication will be in developing awareness of the Charter, how the AI Act and other acts with direct effect protect those rights in practice and how those impact on people's everyday lives. While concrete case law from the AI Act may take a long time to emerge, strong transparency and communication on how fundamental rights are being protected in the public sector use of AI, e.g. through FRIA and public procurement activities, may offer an important source of explanatory stories. In addition, specific public concerns about AI can emerge rapidly based on media attention on the release of a new AI application or a novel type of incident which indicates AI-based harms. A useful communication function may be one that can respond rapidly to a story in the context of the fundamental rights being harmed, the likely acts with direct effect and related competent authorities in play, and an assessment of the uncertainty involved or new questions raised, so as to seek responsive public engagement to the liveness of these issues. In short, raising public literacy in the protection of their fundamental rights is a key component in delivering literacy and critical citizen skills that will underpin trust in AI deployment.
- **Strand 2: A Governance Ecosystem That Promotes Trustworthy AI:** Our recommendations align strongly with the call in this Strand for an agile approach to AI governance and the regulatory framework. In particular we advocate an agile, repeatable and responsive approach to managing the life cycle of regulatory learning mechanisms such as sandboxes and testing in real-world conditions. We advocate that attempts to support voluntary ethical code of practice in the adoption of AI should be clearly seeded from and responsive to the rules and decisions related to protection of fundamental rights for high-risk Ai systems. In this way, self-assessment approaches to ethical AI can be more credibly grounded in the legal decision around the Charter of fundamental rights, the AI Act and other acts with direct effect on protection of rights. Where such decisions are not available, or relevant ethical issues cannot by directly grounded in known legal effect on the development and use of AI, then additional ethical principles or guidelines should be structured as self-administered extensions to know legal fundamental right protections, rather than as an alternative ethical framing, which ultimately will lack a grounding for enforcement as part of the legal framework, and the societal legitimacy that imparts. Further, as we already see open access online tools emerge to support self-assessment or voluntary codes of practice[5] [6] [7], there may be benefit to a common labelling mechanism indexing the capabilities of tools to the potential protection of fundamental rights so that their capabilities can be compared to the 'gold-standard' of legally enforceable fundamental rights risk and mitigation measures. The successful use of UN SDG to label a wide variety of voluntary practices. Similarly, a major component in the implementation of the AI Act is the technical guidance offer by future harmonised standards or common specifications, however existing candidates for adoption in the form of international standards from ISO/IEC JTC1 SC42[8] do not, due to their international nature, have a specific grounding in fundamental rights protections[9]. This current mismatch between technical specification and the Act's focus on fundamental right protection would also benefit from a means of labelling technical measures in standard with their relevance to fundamental rights. Such a mapping from standards to fundamental rights could be integrated into the action laid out in the NSAI AI Standards and Assurance Roadmap[10].

---
[4] https://enterprise.gov.ie/en/publications/national-ai-strategy.html

---

[5] https://regtech.adaptcentre.ie/
[6] https://altai.insight-centre.org/
[7] https://artificialintelligenceact.eu/assessment/eu-ai-act-compliance-checker/
[8] https://www.iso.org/committee/6794475.html

- Strand 3: Driving Adoption of AI in Irish Enterprise: Our recommendations address readiness and the driving of adoption of AI by enterprises by structured mechanisms for communication and regulatory learning to optimise the minimisation of the current legal uncertainties. In addition, proposals for shared data spaces on addressing fundamental rights risk and management information in specific sectors using the facilities of the DGA would offer enterprises a regulated and privacy-aware forum for interacting with societal stakeholder representing groups whose rights need protection in a particular sector..
- Strand 4: AI Serving the Public: Our recommendations prioritise the development of regulatory learning mechanisms targeting the protection of fundamental rights in the public sector use of AI, which feature prominently in the high risk AI categories of Annex III of the Act. This can accelerate the understanding of risks to fundamental rights as public bodies have a direct responsibility to ensure protection of citizens fundamental rights, and are more able to employ high levels of transparency in procurement and compliance actions with peer public bodies, nationally and internationally, than is feasible in competitive commercial markets. Our recommendation of public procurement information data spaces provides a forum for accelerating the sharing of learning from the public sector with regulators, commercial AI providers and societal stakeholders and their representatives. Health, education, public service provision and general public sector functions in employment and cyber security represent possible priorities for initial public sector regulatory learning mechanisms,
- Strand 5: A Strong AI Innovation Ecosystem: Our recommendation prioritises regulator learning in public procurement of high risk AI aa an ideal opportunity to rapidly propagate resolution of legal uncertainties on fundamental rights protection to the private sector. The Irish research and innovation ecosystem offers rich opportunities for collaboration of innovation of AI and its new governance, risk and compliance needs around a set of research centres and spokes, innovation centres and innovation hubs which are thematically aligned in human centric, AI, data analytics, software engineering, networking, data governance, medical devices, learning technology and ICT and AI innovation. Developing the capacity of existing SFI and EI funded centres/hubs as facilitators of engagement between SMEs and start-ups, multinationals, public bodies and societal stakeholder groups may allow them to play an important role in resolving legal uncertainties and developing skills and knowledge capabilities in Ireland and for participation in relevant international R&I collaborations.
- Strand 6: AI Education, Skills and Talent: A successful cooperation between regulatory learning mechanisms and Irish R&I Centres also offers an opportunity to rapidly develop responses to the impact of AI on skills and training. The HEI driving these centres are at the leading edge to grappling with the changes in teaching and learning, especially as advances in generative AI raises profound questions for institutes and their professional accreditation partners about the viability of current teaching and assessment methods of skills and knowledge in fields most likely to be benefits from AI, e.g. engineering, medicine, law, media, creative industries and science. They must also grapple with the use of AI in learning access and assessment itself being a high risk area, in which fundamental rights need to be protected. HEI should be encouraged to place themselves at the forefront internationally in developing new more flexible approaches to learning, both in the use of AI in these different disciplines and the integration of governance, risk and compliance requirements into these new AI-based practices. If this leads to a robust and viable HEI sector, Ireland will be much better placed to build AI skills, attract the best international talent and ensure equality and diversity as AI becomes widely adopted in the workplace.
- Strand 7: A Supportive and Secure Infrastructure for AI: As with national digital infrastructure, robust skills and capacity in securing AI systems in the public and private sector will be a critical prerequisite to securing public trust in those systems. This is of particular relevance in Ireland nationally due to the disproportionally high distribution of data centres in Ireland, where international MNC serve the Single Market.

---

[9] https://publications.jrc.ec.europa.eu/repository/handle/JRC132833
[10] https://www.nsai.ie/images/uploads/general/NSAI_AI_report_digital.pdf